\documentstyle[twocolumn,eqsecnum,psfig,aps]{revtex}

\begin{document}
\draft
\title{Transfer Across Random versus Deterministic Fractal Interfaces.\\
}
\author{M. Filoche$^{1}$ and B. Sapoval$^{1,2}$}
\address{$^{1}$Laboratoire de Physique de la Mati\`{e}re Condens\'{e}e,\\
C.N.R.S. Ecole Polytechnique, 91128 Palaiseau, France \\
$^{2}$Centre de Math\'{e}matiques et de leurs Applications,\\
Ecole Normale Sup\'{e}rieure, 94140 Cachan, France}
\date{\today }
\maketitle

\begin{abstract}
A numerical study of the transfer across random fractal surfaces shows that
their response are very close to the response of deterministic model
geometries with the same fractal dimension. The simulations of several
interfaces with prefractal geometries show that, within very good
approximation, the flux depends only on a few characteristic features of the
interface geometry: the lower and higher cut-offs and the fractal dimension.
Although the active zones are different for different geometries, the
electrode responses are very nearly the same. In that sense, the fractal
dimension is the essential ''universal'' exponent which determines the net
transfer.
\end{abstract}

\pacs{{\bf PACS:} 41.20.Cv - 82.65.Jv - 61.43.Hv}

Many random processes such as aggregation, diffusion, fracture and
percolation, build fractal objects \cite{Vicsek,Meakin}. Fractal geometry
essentially describes hierarchical structures \cite{Mandel}. If properties
of these random systems depend on the hierarchical character of their
geometry, then the study of a deterministic structure with the same fractal
dimension may provide a good approximation of the random system properties 
\cite{sapfrac}. The question is significant since fractal and pre-fractal
geometries are widely used in mathematical approaches or numerical
simulations as a convenient model of irregularity. They are also more simply
addressed by algebraic calculations and incorporated into numerical models
for computer simulation. It is then an important matter to decide whether
simple deterministic, {\it artificial,} fractals could help determine the
properties of random, {\it natural,} fractals~\cite{MandelGiven,deArcangelis}%
. In particular, it is a question whether experiments performed on model
fractal geometries \cite{Sapoval96} may help understand the behavior of real
complex structures.

The property which is discussed here is the Laplacian transport to and
across irregular and fractal interfaces. Such transport phenomena\ are often
encountered in nature or in technical processes: properties of rough
electrodes in electrochemistry, steady-state diffusion towards irregular
membranes in physiological processes, the Eley-Rideal mechanism in
heterogeneous catalysis in porous catalysts, and in NMR relaxation in porous
media. In each of these examples, the interface presents a {\it finite}
transfer rate, like a redox reaction, or a finite permeability, or reaction
rate which is due to specific physical or chemical processes.

\begin{figure}
\centerline{
\psfig{file=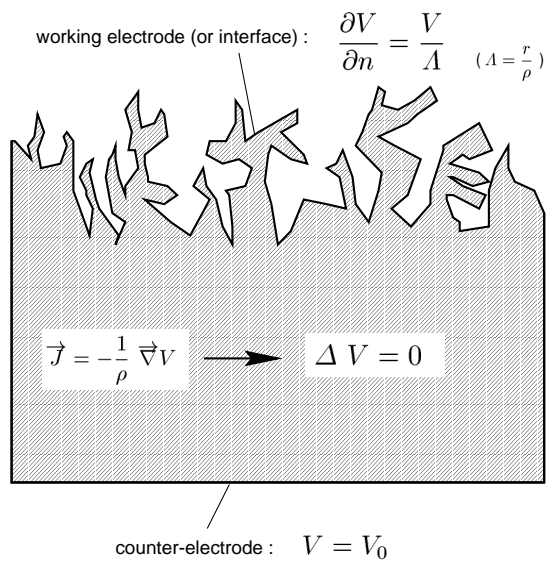,width=.4\textwidth}}
\smallskip
\caption{Schematic representation of an electrochemical cell.}
\label{fig:cell}
\end{figure}

The mathematical formulation of the problem is simple. One considers the
current flowing through an electrochemical cell as shown in Fig.~\ref
{fig:cell}. The current $\vec{J}$ is proportional to the Laplacian field $%
\vec{\nabla}V$, which can be viewed as an electrostatic field in
electrochemistry, or a particle concentration field in diffusion problems.
Then the flux and field are related by classical equations of the type $\vec{%
J}=-\sigma \vec{\nabla}V$, where $\sigma $ is the electrolyte conductivity
(or particle diffusivity in diffusion or heterogeneous catalysis). The
conservation of this current throughout the bulk yields the Laplace equation
for the potential $V$:

\begin{equation}
div(-\sigma ~\vec{\nabla}V)=0\qquad \Rightarrow \qquad \Delta ~V~=~0
\end{equation}

The boundary presents a finite resistance to the current flow. In the
simplest case, this resistance can be expressed by a linear relation linking
the current density across the boundary to the potential drop across that
boundary. The local flux and potential drop are then linked by transport
coefficients, like the faradaic resistance in electrochemistry, the membrane
permeability in physiological processes, or again the surface reactivity in
catalysis. If one assumes that the outside of the irregular boundary is at
zero potential, current conservation at the boundary leads to the following
relation:

\begin{equation}
\vec{J}\cdot \vec{n}~=~-\frac{V}{r}
\end{equation}
or 
\begin{equation}
\frac{\partial V}{\partial n}~=~\frac{V}{\Lambda }\qquad {\rm with}\qquad 
{\it \Lambda =}~\sigma r
\end{equation}

The parameter $\Lambda $ is homogeneous to a length. Given the geometry, the
value of this parameter determines the behavior of the system \cite
{Sapo94,Sapoval99}. The overall response of such a system is measured by one
scalar quantity, its impedance $Z_{tot}$, which is the ratio between the
applied potential and the total flux :

\begin{equation}
Z_{tot}~=~\frac{V}{\Phi}
\end{equation}

The contribution of the finite interface resistivity to this global
impedance is given by a ``spectroscopic'' impedance, defined as:~$%
Z_{spec.}~=~Z_{tot}-Z_{0}$,~ $Z_{0}$ being the impedance of the cell with
zero interface resistivity \cite{Sapoval99}. The main result discussed below
is that the electrode impedance $Z_{spec.}$, is nearly independent of the
random character of the fractal interface, even though the regions where the
current is concentrated are very different. This is found from a numerical
comparison between impedances of deterministic and random electrodes with
the same fractal dimension. Two cases are studied: (a) deterministic and
random von Koch electrodes (dimension $D_{f}=\ln 4/\ln 3),$ (b) a
deterministic electrode of dimension $D_{f}=4/3$ and a self-avoiding random
walk geometry with the same dimension.

\begin{figure}
\centerline{
\psfig{file=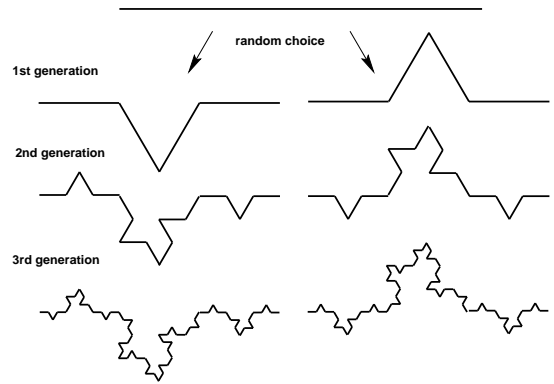,width=.4\textwidth}}
\smallskip
\caption{The building process of random von Koch curves. The same
random process can create various interface topographies. They share
the same size, the same perimeter, and the same fractal dimension.}
\label{fig:VonKochrand}
\end{figure}

\begin{figure}
\centerline{
\psfig{file=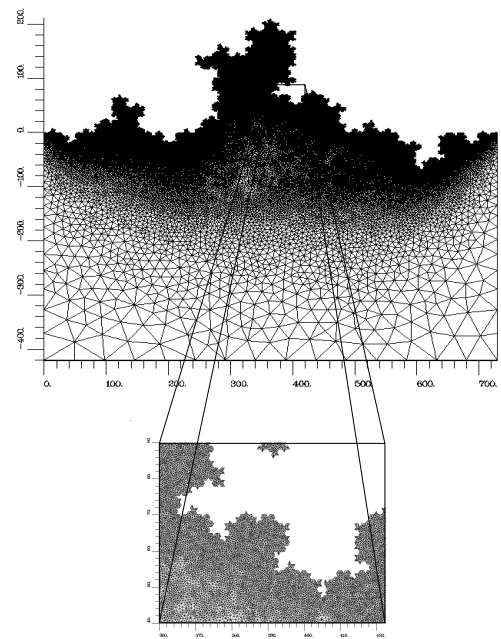,width=.4\textwidth}}
\smallskip
\caption{A finite element mesh for the $6^{th}$ generation von Koch
random electrode. Top:~an example of a finite element mesh for the
$6^{th}$ generation interface. Bottom:~local zoom of the mesh.}
\label{fig:meshes}
\end{figure}

The deterministic von Koch curve, or classical snowflake curve, is
obtained by dividing a line segment in three equal parts, removing the
central segment and replacing it by two other identical segments which
form an equilateral triangle. A random von Koch curve can be defined
simply by choosing randomly the side of the segment where the triangle
is created at each step of the building process. This is shown in
Fig.~\ref{fig:VonKochrand}. After three or more generations, it looks
more like a realistic random boundary than a simple mathematical
curve. It is then possible to automatically generate different
boundaries that have the same fractal dimension and the same
perimeter. By definition fractal geometries exhibit a large scale of
lengths. For instance, at the sixth generation, the ratio between the
smallest feature $l$ (smaller cut-off) of the irregular boundary and
the diameter $L$ (larger cut-off) is $3^{6}=729$ while the length of
the perimeter is $L_{p}=4^{6}l=4096l$. Computing on a regular grid
within such geometries would be very memory and time-consuming. A
finite element method is then used. The standard variational
formulation of the problem is discretized with a triangular mesh,
obtained from a Delaunay-Vorono{\"{\i }} tessellation and
$P_{1}$-Lagrange interpolation.  The linear system obtained in such a
way is solved by using the Cholesky method, from the Finite Element
Library MODULEF \cite{Modulef}. Examples of meshes with a $6^{{\rm
th}}$ generation boundary are shown in Fig.~\ref{fig:meshes}.

\begin{figure}
\centerline{
\psfig{file=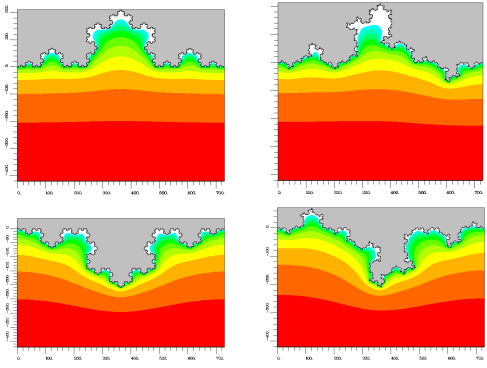,width=.4\textwidth}}
\smallskip
\caption{Isopotential curves for von Koch deterministic and random
electrodes with $\Lambda =0$ (Dirichlet boundary condition). The
equipotential lines are the lines separating regions of exponentially
decreasing potential: V=1 at the bottom then 1/2, 1/4, 1/8,... The
current density is proportional to the gradient of the potential. The
current is then large in regions where the curves are close. Note that
the current flows through the interface primarily at the tips. These
active zones are found at very different locations for different
electrodes.}
\label{fig:isocurves}
\end{figure}

Computations were carried out for the two deterministic boundary
geometries and the two random geometries of generation 6 shown in
Fig.~\ref{fig:isocurves}. The figure presents the isocurves of the
potential for $\Lambda = 0$. Since the current density is proportional
to the gradient of the potential, one can detect regions of large
current density from the distance between two consecutive isocurves:
the closer the equipotentials, the larger the current density. As
expected, most of the current flow through the irregular interface at
the tips. This gives a very different current map for each
geometry. Therefore, for the different electrodes the active zones are
very different.

\begin{figure}
\centerline{
\psfig{file=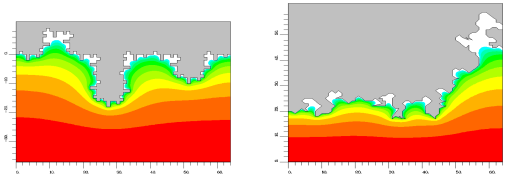,width=.4\textwidth}}
\smallskip
\caption{Isopotential curves for deterministic and random electrodes
of fractal dimension 4/3 with $\Lambda =0$ (Dirichlet boundary
condition). Same color code as Fig. \ref{fig:isocurves}. The active
zones are entirely different.}
\label{fig:4/3}
\end{figure}

The second type of electrodes to be compared is shown in
Fig.~\ref{fig:4/3}.  The top figure shows the second generation of a
deterministic fractal electrode with dimension $D_{f}=\ln 16/\ln
8=4/3$ while the bottom represents a particular self-avoiding walk
with the same 4/3 fractal dimension. Both electrodes have the same
perimeter and the same smaller cut-off. Here, even more than above,
the active zones are totally different.

\begin{figure}
\centerline{
\psfig{file=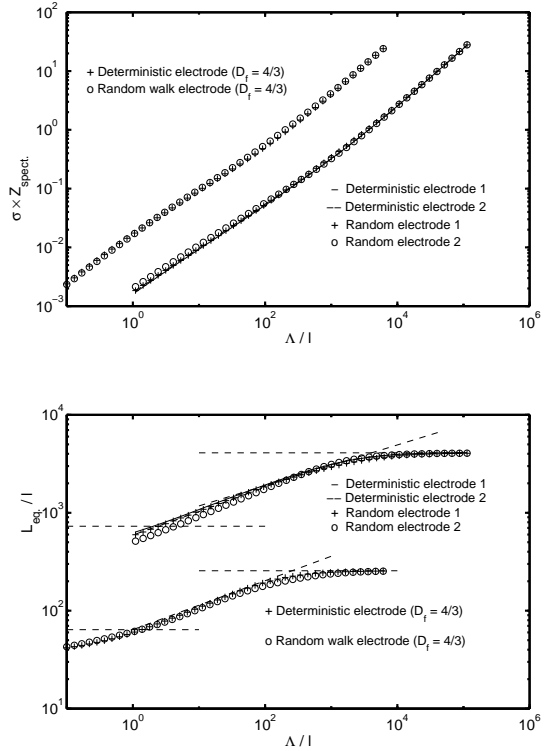,width=.4\textwidth}}
\smallskip
\caption{Top: Plots of the electrode impedance $Z_{spec.}$ as a function of $%
\Lambda /l=r\sigma /l$, for various deterministic and random geometries.
Note curve similarities despite very different current maps. Bottom: Plots
of the ''equivalent length'' of the working surfaces defined by $%
L_{eq.}=r/Z_{spec.}$. Approximate expression of $Z_{spec.}$ and $L_{eq.}$
mentioned in the text, are indicated by the dashed lines.}
\label{fig:impedances}
\end{figure}

For each geometry, the impedances have been computed for an extended
range of the surface resistivity $r$. The results are shown in
Fig.~\ref{fig:impedances} for two categories of geometries : $6^{{\rm
th}}$ generation of von Koch electrodes and the two electrodes of
Fig.~\ref{fig:4/3}. The parameter ${\it \Lambda /}l=$ $\sigma r{\it
/}l$ ranges between $1$ and 10$^{5}$ for generation 6 and between
10$^{-1}$ and $5.10^{3} $ for the second type. The limitation of the
range is due to limitations in computer time and memory.

It is striking that, despite very different current distribution in the bulk
and at the interface, the impedances are very close for all values of the
surface resistivity. The behavior of different interfaces are nearly
indistinguishable: random and deterministic interfaces behave in the same
manner. This could be considered as a partial answer to the question{\bf \ }%
''Can One Hear the Shape of an Electrode?'', addressed in \cite
{Sapoval99,Filoche99}{\bf . }In this frame, the main parameters drawn from
practical impedance spectroscopy measurements would only be the size, the
perimeter and the equivalent fractal dimension of the interface.

A more demanding comparison between the impedances can be made by comparing
the values of $r/Z$ as shown in Fig.~\ref{fig:impedances}. This quantity can
be identified as an equivalent active length $L_{eq.}$ \cite{note1}. One
finds three successives regimes, ${\it \Lambda <}l$; $l{\it <\Lambda <}$ $%
L_{p,}$ and finally $L_{p}{\it <\Lambda }$ separated by smooth crossovers.
These regimes can easily be compared to the so-called ''land surveyor
approximation'' \cite{Filoche97}. This method allows one to compute $%
Z_{spec.}$ through a finite size renormalization of the interface geometry,
without solving the Laplace equation. For small $r$ (or ${\it \Lambda <<1}$)
there is a linear regime in which $Z_{spec.}$ is proportional to $r$, that
is $Z_{spec.}=r/L_{eq.}$ with $L_{eq.}\approx L$~\cite{Sapoval99}. For
values of ${\it \Lambda }$\mbox{$>$} $l$ there is a fractal regime in which,
in first approximation, $Z_{spec.}=(r/\Lambda {\it )}(l/L){\it (\Lambda /}%
l)^{1/D_{f}}$ and $L_{eq.}=$ $L({\it \Lambda }/l)^{(D_{f}-1)/D_{f}}$ (for
more detailed expressions of the exponents, see \cite{Halsey,Ball93,Ruiz}).
Finally, for values of ${\it \Lambda }$ much larger than the perimeter
length $L_{p}$, the exact value is $Z_{spec.}=r/L_{p}$ and $L_{eq.}=$
\thinspace $L_{p}$. These three asymptotic behaviors are shown in the figure
and are found to match the numerical results with good accuracy.

Note that the electrodes of Fig.~\ref{fig:4/3} are in some sense ''poor''
fractals because the range of geometrical scaling is relatively small and it
has been a matter of debate recently whether the fractal concept should be
of any use when the scaling range of the geometry is too small. For the
phenomena considered here, one can observe that the fractal description of
this limited range geometry is really useful.

In summary, one has shown on several examples that the net transfer across
an irregular surface is nearly independent of the randomness of its
geometry, although it depends strongly on the geometry through its fractal
dimension. The fact that the overall response remains the same indicates
that, buried in the fractal description, there exist the geometrical
correlations that govern the overall effect of screening {\it at different
scales}. In that sense the response is ''universal'' within a very good
approximation for the category of curves considered here.

The authors wish to acknowledge useful discussions with P.~Jones and
N.~Makarov. This research was supported by N.A.T.O. grant C.R.G.900483. The
Laboratoire de Physique de la Mati\`{e}re Condens\'{e}e is ``Unit\'{e} Mixte
de Recherches du Centre National de la Recherche Scientifique No.~7643''.

% REFERENCES


\begin{references}
\bibitem{note1}  Note that this length is different from the length of the
active zone as defined in reference \cite{Sapoval99,Filoche99}.
\end{references}
\end{document}